\documentclass[pra,twocolumn,showpacs,preprintnumbers,amsmath,
amssymb,floatfix]{revtex4}
\usepackage{hyperref}
\usepackage{epsfig}
\newcommand{\beq}{\begin{equation}}
\newcommand{\eeq}{\end{equation}}
\newcommand{\beqa}{\begin{eqnarray}}
\newcommand{\eeqa}{\end{eqnarray}}
\newcommand{\ba}{\begin{array}}
\newcommand{\ea}{\end{array}}
\begin{document}

\title{Bulk and Collective Properties of a Dilute Fermi Gas 
in the BCS-BEC Crossover} 
\author{N. Manini and L. Salasnich} 
\affiliation{Dipartimento di Fisica, Universit\`a di Milano, \\
Istituto Nazionale per la Fisica della Materia, 
Unit\`a di Milano, \\ 
Via Celoria 16, 20133 Milano, Italy}

\begin{abstract} 
We investigate the zero-temperature properties of a dilute 
two-component Fermi gas with attractive 
interspecies interaction in the BCS-BEC crossover. 
We build an efficient parametrization of the energy per particle based on
Monte Carlo data and asymptotic behavior.
This parametrization provides, in turn, analytical expressions for several
bulk properties of the system such as the chemical potential, the pressure
and the sound velocity.
%
%
In addition, by using a time-dependent density functional approach, 
we determine the collective modes of the Fermi gas under harmonic 
confinement. The calculated collective frequencies are compared 
to experimental data on confined vapors of $^6$Li atoms 
and with other theoretical predictions. 
\end{abstract}


\maketitle

\section{Introduction}

A hot topic in nowadays' many-body physics is the study of Fermi gases at
ultra-low temperature. Indeed, current experiments with atomic vapors are
able to operate in the regime of deep Fermi degeneracy \cite{p1,p2}.
The two-component Fermi gases of these experiments
are dilute because the effective range $R_0$ 
of the interaction is much smaller than the mean interparticle 
distance, i.e. $k_F R_0 \ll 1$ where $k_F=(3\pi^2 n)^{1/3}$ 
is the Fermi wave vector and $n$ is the gas number density. 
Even in this dilute regime the s-wave scattering length $a$ 
can be very large: the interaction parameter $k_F a$ 
can be varied over a very wide range using the 
Feshbach resonance technique, which permits to vary the magnitude and 
the sign of $a$. The available experimental 
data on $^{6}$Li atoms are concentrated across the resonance, 
where $a$ goes from large negative to large positive values \cite{p3,p4} 
and where a crossover from a Bardeen-Cooper-Schrieffer (BCS) superfluid 
to a Bose-Einstein condensate (BEC) of molecular 
pairs has been predicted \cite{p5,p6,p7}. 

In this paper we propose a reliable analytical fitting formula 
for the energy per particle of a homogeneous two-component 
Fermi gas, by analyzing the fixed-node Monte Carlo 
data of Astrakharchik {\it et al.}\ \cite{p8}. From this analytical 
formula it is straightforward to calculate several bulk properties 
of the system by means of standard themodynamical relations. 
This fitting formula enables us to calculate also the collective 
modes of the Fermi gas under harmonic confinement by 
using the hydrodynamic theory in the local-density 
approximation (LDA), including also a quantum pressure term. 
We compare our results with other theoretical 
predictions \cite{p9,p10,p11,p12,p13,p14} and, in particular, 
with the experimental frequencies of the collective 
breathing modes \cite{p3,p4}. 

\section{Bulk properties}

At zero temperature, the bulk energy per particle $\cal E$ 
of a dilute Fermi gas can be written as 
\beq 
{\cal E} = {3\over 5} \epsilon_F \; \epsilon(x) \; , 
\eeq 
where $\epsilon_F = \hbar^2k_F^2/(2m)$ is the Fermi energy 
and $\epsilon(x)$ is a yet unknown function of the interaction 
parameter $x=k_Fa$. 
In the weakly attractive regime ($-1\ll x<0$) 
one expects a BCS Fermi gas 
of weakly bound Cooper pairs.
As the superfluid gap correction 
is exponentially small, the function $\epsilon(x)$ should follow
the Fermi-gas expansion \cite{p15}  
\beq 
\epsilon(x)=1+{10\over 9\pi} x + 
{4(11 - 2\ln{(2)}) \over 21\pi^2} x^2 + ... \; .  
\eeq 
In the weak BEC regime ($0<x\ll 1$) 
one expects a weakly repulsive Bose gas of dimers. 
Such Bose-condensed molecules of mass $m_M=2m$ and
density $n_M=n/2$  interact with a positive scattering length 
$a_M=0.6 a$, as predicted by Petrov {\it et al.}\ \cite{p16}. 
In this regime, 
after subtraction of the molecular binding energy, the function 
$\epsilon(x)$ should follow the Bose-gas expansion \cite{p17} 
\beq 
\epsilon(x) = {5\over 18 \pi} {a_M\over a} x 
\left[ 1 + {128\over 15\sqrt{6\pi^2}} 
\left({a_M\over a}\right)^{3/2} x^{3/2} + ... \right] \; .  
\eeq 
In the so-called unitarity limit ($x=\pm \infty$) one expects that 
the energy per particle is proportional to that of a non-interacting 
Fermi gas \cite{p18}. The fixed-node diffusion Monte-Carlo calculation 
of Astrakharchik {\it et al.} \cite{p8} finds 
\beq 
\epsilon(x=\pm \infty) = 0.42 \pm 0.01 \; , 
\eeq 
while an analogous calculaton of Carlson {\it et al.} \cite{p19} gave 
$\epsilon(x=\pm \infty) = 0.44 \pm 0.01$. 
The calculation of Astrakharchik {\it et al.} \cite{p8} 
is quite complete and gives the behavior of the energy of system across 
the unitarity limit.
It is a standard convention to use the inverse interaction parameter
$y=1/x=1/(k_Fa)$ as the independent variable.
In Fig.~1 we plot the data of $\epsilon(y)$ reported 
by Astrakharchik {\it et al.}\ \cite{p8,p20}. 
On the basis of the data of Carlson {\it et al.}\ \cite{p19}, Bulgac and
Bertsch \cite{p14} proposed the following behavior of $\epsilon(y)$ near
$y=0$:
\beq  
\epsilon(y) = \xi - \zeta \; y + ... 
\eeq
with $\xi =0.44$ and $\zeta = 1$ for both positive and negative $y$.
The denser data of Ref.~\cite{p8} suggest instead a continuous but
not differentiable behavior of $\epsilon(y)$ near $y=0$, namely with
$\xi=0.42$ and $\zeta =\zeta_- = 1$ in the BCS region ($y<0$) but $\zeta
 =\zeta_+=1/3$ in the BEC region ($y>0$).
As expected, for large $|y|$, the Monte-Carlo data shown in Fig.~1 
follow the asymptotic trends of Eq.~(1) and Eq.~(2).

We propose here the following analytical fitting formula
\beq 
\epsilon(y) = \alpha_1 - \alpha_2 
\arctan{\left( \alpha_3 \; y \; 
{\beta_1 + |y| \over \beta_2 + |y|} \right)}  \; , 
\eeq
interpolating the Monte Carlo energy per particle and the limiting
behaviors for large and small $|y|$.
Here the parameter $\alpha_1$ is fixed by the value $\xi$ of $\epsilon(y)$
at $y=0$, the parameter $\alpha_2$ is fixed by the value of $\epsilon(y)$
at $y=\infty$, and $\alpha_3$ is fixed by the asymptotic $1/y$ coefficient
of $\epsilon(y)$ at large $|y|$ (Eqs.~(2) and (3)).
The ratio $\beta_1/\beta_2$ is determined by the linear behavior $\zeta$
of $\epsilon(y)$ near $y=0$.
The value of $\beta_1$ is then determined by minimizing the mean square
deviation from the Monte-Carlo data \cite{p21}. 
Of course, we consider two different set of parameters: one set in the BCS
region ($y<0$) and a separate set in the BEC region ($y>0$). Table 1 
reports the values of these parameters.

\begin{table}[]
\begin{center}
\renewcommand{\arraystretch}{1.3}
\begin{tabular}{cclcl}
\hline
\hline
 	&	\multicolumn{2}{c}{BCS	($y<0$)}	&	\multicolumn{2}{c}{BEC	($y>0$)}	\\
	& expression & value& expression & value\\
\hline
$\alpha_1$	&	$\xi$	&	0.4200	&	$\xi$	&	0.4200	\\
$\alpha_2$	&	$\frac 2\pi(1-\alpha_1)$	&	0.3692	&	$\frac 2\pi\alpha_1$	&	0.2674	\\
$\alpha_3$	&	$\frac{9\pi}{10}\alpha_2$	&	1.0440	&	$\frac{18\pi}{5}\alpha_2\frac{a_M}{a}$	&	5.0400	\\
$\beta_1$	&	[fitted]	&	1.4328	&	[fitted]	&	0.1126	\\
$\beta_2$	&	$\alpha_2\alpha_3\beta_1/\zeta_-$	&	0.5523	&	$\alpha_2\alpha_3\beta_1/\zeta_+$	&	0.4552	\\
\hline
\hline
\end{tabular}
\renewcommand{\arraystretch}{1}
\end{center}
\caption{\label{param-table}
Parameters of the fitting function (6).
}
\end{table}

Figure 1 compares this fitting function (solid curve) to the Monte Carlo
data. For the sake of completeness, in Fig.~1 we also show the dotted curve
obtained with the [2,2] Pad\'e approximation of Kim and Zubarev \cite{p11},
based only on the asymptotes and the Monte-Carlo value \cite{p19} at $y=0$. 
Our parametric formula is more accurate, especially around $y=0$.

\begin{figure}
\centerline{
\epsfig{file=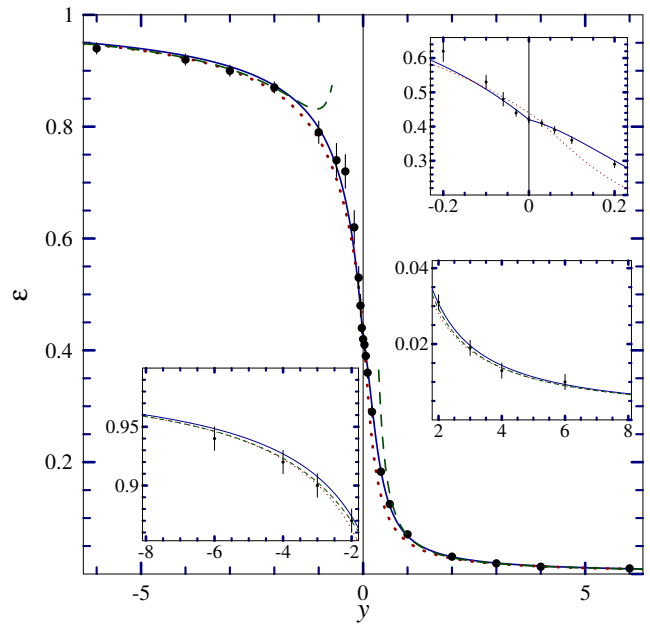,width=85mm,clip=}}
\caption{\label{energy:fig}
The energy per particle $\epsilon(y)$ as defined in Eq.~(1), 
where $y=1/x=1/(k_Fa)$. 
Solid circles represent the fixed-node Monte-Carlo data of
Ref.~\cite{p8}.
Solid line is the parametric function (6) based 
on the values of Table 1. 
Dotted line is the Pad\'e approximation of Ref.~\cite{p11}.
Dashed lines represent the asymptotic expressions Eq.~(2) and Eq.\ (3). }
\end{figure}

\begin{figure}
\centerline{\epsfig{file=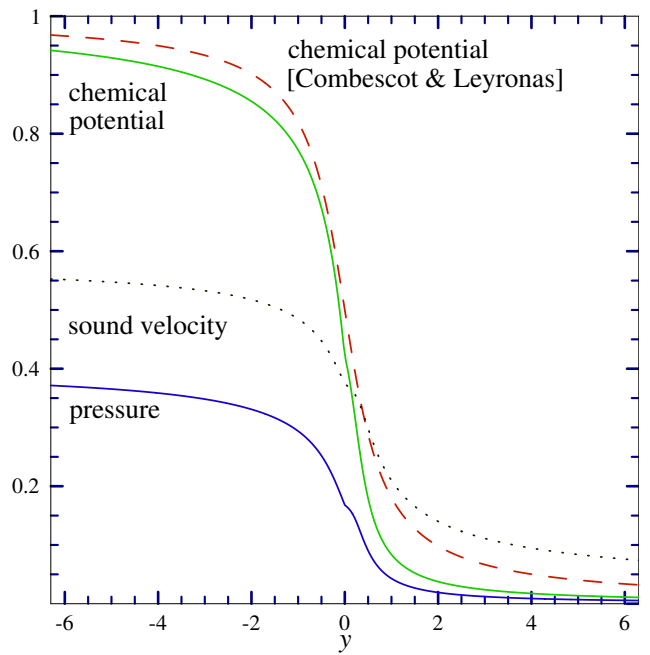,width=85mm,clip=}}
\caption{\label{thermodynamics:fig}
Chemical potential $\mu/\epsilon_F$, pressure $P/(n\epsilon_F)$ and sound
velocity $c_s/v_F$,
obtained from the parametric function (6).
Dashed line represents the simple model
$\mu/\epsilon_F=\frac 12 - \frac 1{\pi} \arctan(\pi y/2)$
of Ref.~\cite{p12}.
}
\end{figure}

The advantage of a functional parametrization of $\epsilon(y)$ 
is that it allows straightforward analytical calculations of 
several ground-state physical properties 
of the bulk Fermi gas \cite{p22}. 
For example, the chemical potential $\mu$ is given by 
\beq 
\mu = {\partial (n{\cal E}) \over \partial n} = 
\epsilon_F \left( \epsilon(y) - {y \over 5} \epsilon'(y) \right) \; , 
\eeq 
as found by using Eq.~(1) and taking into account 
that $\partial y/\partial n = - y/(3n)$, while 
the pressure $P$ reads 
\beq 
P= n^2 {\partial {\cal E}\over \partial n} = 
n \; \epsilon_F \left( {2\over 5} \epsilon(y) - 
{y \over 5} \epsilon'(y) \right) \; . 
\eeq 
The sound velocity $c_s$ is instead obtained as 
$c_s^2 = (n/m)\partial \mu /\partial n$, from which we get 
\beq 
c_s = v_F \; \sqrt{ {1 \over 3} \epsilon(y) - {y \over 5} 
\epsilon'(y) + {y^2 \over 30}\epsilon''(y) } \; , 
\eeq 
where $v_F= (2\epsilon_F/m)^{1/2}$ is the Fermi velocity. 
Figure 2 reports the chemical potential $\mu$, the pressure $P$ and the
sound velocity $c_s$ as a function of $y$.
Our theory predicts that all these macroscopic properties show a kink at the
unitarity point, due to $\zeta_- \neq \zeta_+$.
Figure 2 shows also the curve of the chemical potential obtained with the
simple analytical model proposed by Combescot and Leyronas \cite{p12}.
The sound velocity $c_s$ is accessible experimentally, and the dotted curve
of Fig.~2 is our prediction of the way $c_s$ evolves from $v_F/\sqrt{3}$ to
zero through the BCS-BEC crossover.

\section{Harmonically confined gas}
 
We consider now the effect of confinement due to an external 
anisotropic harmonic potential 
\beq 
U({\bf r})={m\over 2}\left(\omega_{\rho}^2 \rho^2 + 
\omega_z^2 z^2\right) \; , 
\eeq
where $\omega_{\rho}$ is the cylindric 
radial frequency and $\omega_z$ is the cylindric 
longitudinal frequency. Assuming that the density field 
$n({\bf r},t)$ varies sufficiently slowly (this 
assumption is at the basis of the LDA), 
at each point ${\bf r}$ the gas can be considered in local 
equilibrium, and the local chemical potential is $\mu[n({\bf r},t)]$. 
Within the LDA, the dynamics can be described by means of 
the hydrodynamic equations of superfluids 
\beqa
{\partial n \over \partial t} 
+ {\bf \nabla} \cdot (n {\bf v}) &=& 0 \; , 
\\
m {\partial n \over \partial t} + 
\nabla \left( \mu[n({\bf r}, t)] + U({\bf r}) + 
{1\over 2} m v^2 \right) &=& 0 \; , 
\eeqa
where ${\bf v}({\bf r},t)$ is the velocity field,
and $\mu[n]$ is the chemical potential of Eq.~(7).
It has been shown by Cozzini and Stringari \cite{p23} 
that assuming a power-law dependence $\mu = \mu_0 \; n^{\gamma}$ 
for the chemical potential (polytropic equation of state 
\cite{p12}) from Eqs. (11-12) one finds analytic 
expressions for the collective 
frequencies.
In particular, for the very elongated cigar-shaped traps used in recent
experiments ($\omega_r/\omega_z >20$), the collective radial breathing
mode frequency $\Omega_{\rho}$ is given by \cite{p23}
\beq 
\Omega_{\rho} = \sqrt{2(\gamma + 1)} \; \omega_{\rho} \; , 
\eeq
while the collective longitudinal breathing mode $\Omega_{z}$ is 
\beq
\Omega_z = \sqrt{3\gamma +2 \over \gamma +1} \; \omega_z \; .  
\eeq

\begin{figure}
\centerline{\epsfig{file=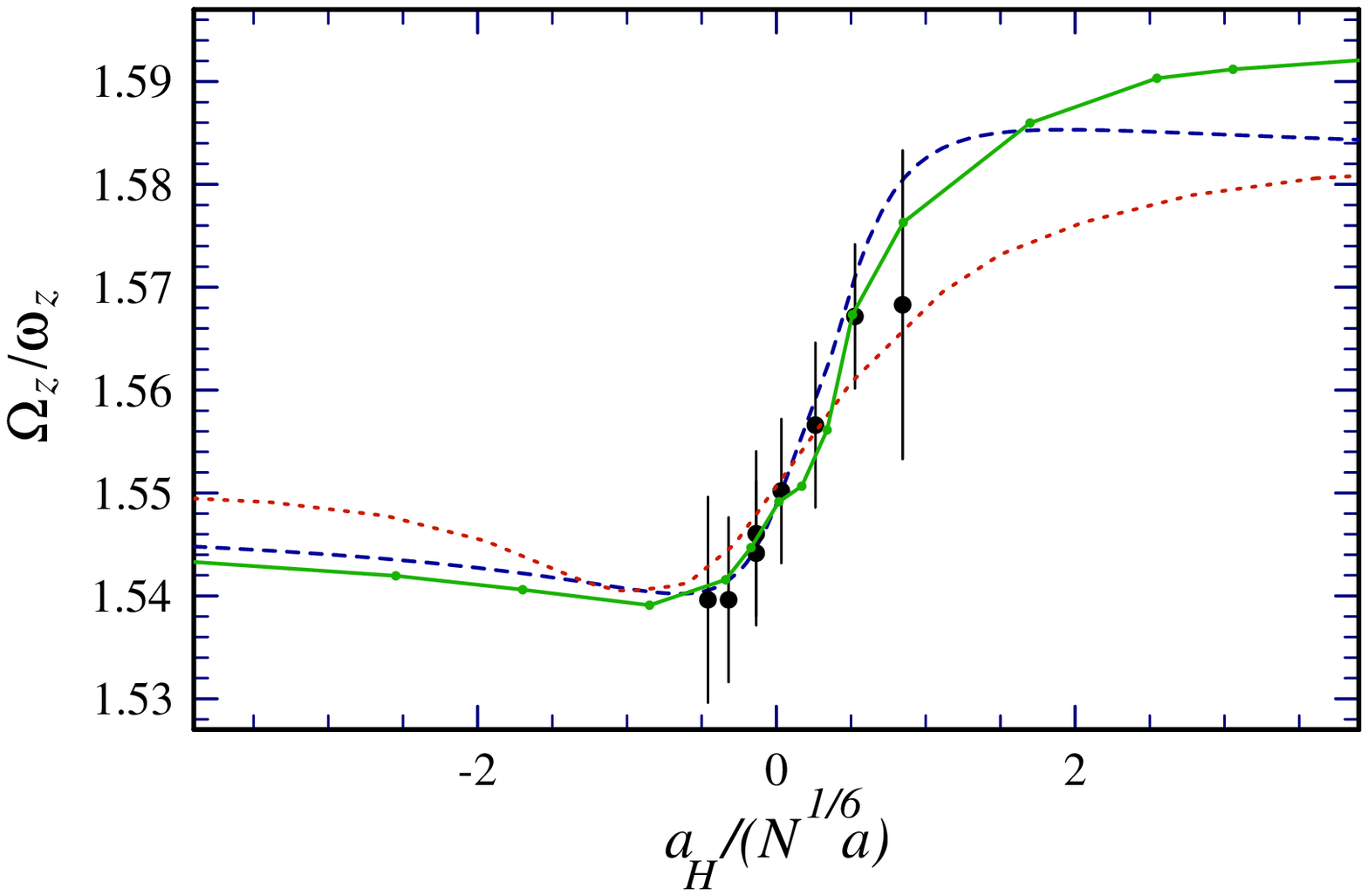,width=85mm,clip=}}
\centerline{\epsfig{file=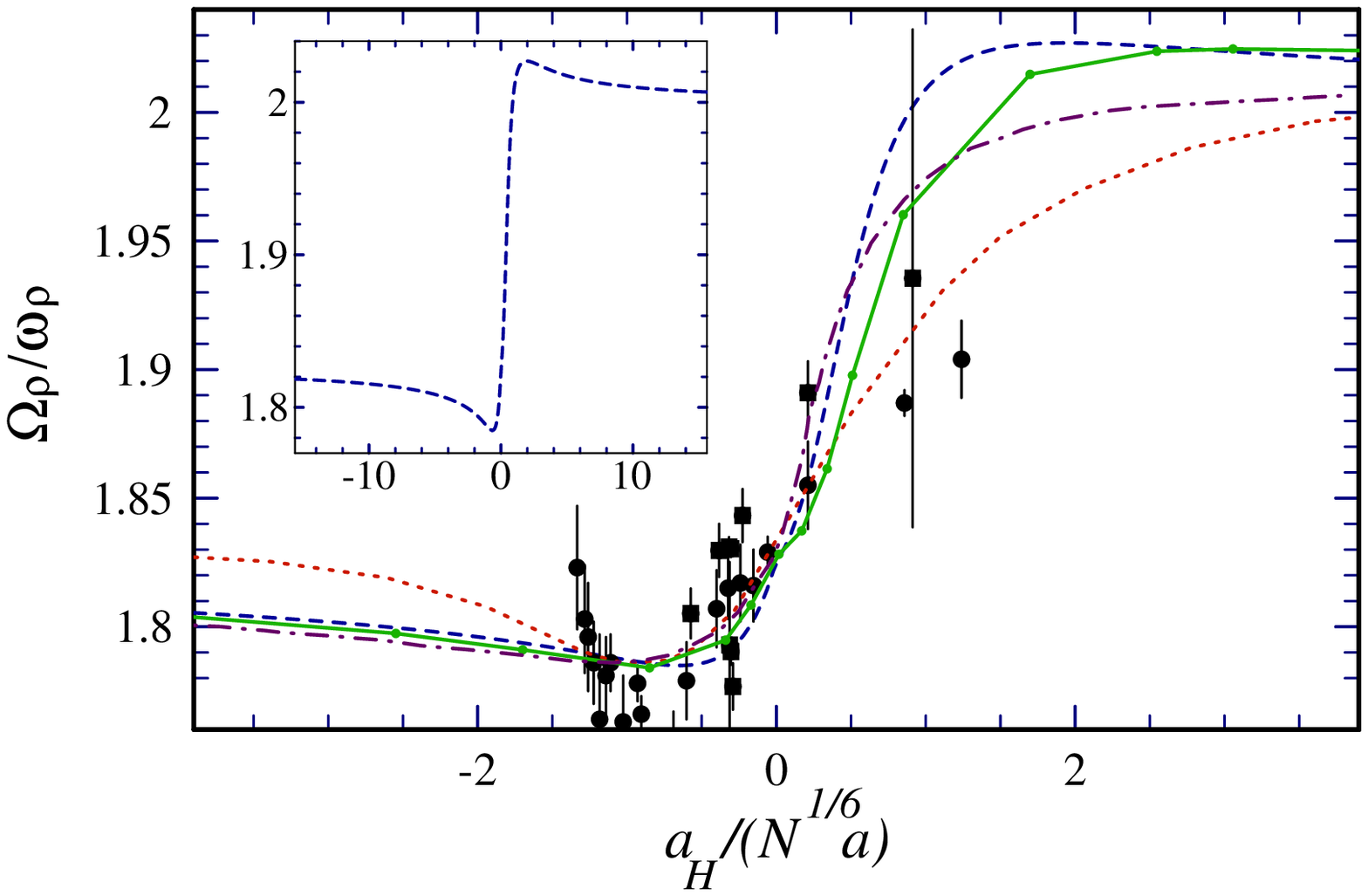,width=85mm,clip=}}
\caption{\label{omega:fig}
Frequencies of the longitudinal and radial collective breathing modes of
the 2-component Fermi gas as a function of 
$a_H/(N^{1/6} a) = 2^{1/2} 3^{1/6}/(k_F a)=2^{1/2} 3^{1/6} y$.
$\Omega_z$ experimental data from Ref.~\cite{p4}, and $\Omega_{\rho}$
experimental data from Ref.~\cite{p3}a (squares) and Ref.~\cite{p3}b
(circles), both for $^6$Li.
Joined dots with solid lines: TDDFT of Eq.~(16), with $N=4\times 10^5$,
$\omega_\rho/\omega_z = 31.9$ for $\Omega_z$ and  with $N=2\times 10^5$,
$\omega_\rho/\omega_z = 22.1$ for $\Omega_\rho$.
Dot-dashed line: variational approximation to TDDFT, Ref.~\cite{p11}.
Dashed line: polytropic hydrodynamic approximation Eq. (13-14).
Dotted line: the mean-field BCS of Ref.~\cite{p10}.
}
\end{figure}

In our problem we introduce an effective polytropic index 
$\gamma$ as the logarithmic derivative of the chemical 
potential $\mu$, that is 
\beq 
\gamma = {n \over \mu} {\partial \mu \over \partial n} = 
{  {2 \over 3} \epsilon(y) - 
{2y \over 5} \epsilon'(y) + {y^2 \over 15}\epsilon''(y) 
\over \epsilon(y) - {y \over 5} \epsilon'(y) } \; . 
\eeq 
We have verified that indeed $\gamma$ remains relatively close to unity for
all $y$: the results of the local polytropic equation are thus useful 
to have a simple analytical prediction of the collective frequencies. 
Based on this polytropic hydrodynamic approximation (PHA), by using
Eq.~(6) we obtain the breathing-mode frequencies shown in Fig.~3 as dashed
lines.

The analytical prediction of Eqs.~(13-15) can be improved by releasing the
polytropic approximation and explicitly integrating Eqs.~(11-12).
We have done such a calculation by including also a quantum pressure term
$(-\hbar^2 \nabla^2 \sqrt{n})/(2m\sqrt{n})$ in Eq.~(12).
In practice, one must solve the following 
time-dependent nonlinear Schr\"odinger equation 
\beq 
i\hbar {\partial \over \partial t} \psi({\bf r},t) 
= \left[ -{\hbar^2 \over 2m} \nabla^2 
+ U({\bf r}) + \mu[n({\bf r}, t)] \right] \psi({\bf r},t) \; , 
\eeq
where $\psi({\bf r},t)$ is the superfluid wave function such that $n({\bf
r},t) = |\psi({\bf r},t)|^2$, ${\bf v} = \hbar \, \nabla
\ln(\psi/n^{1/2})/(i m)$, and $\mu[n]$ is the chemical potential of
Eq.~(7).
Equation~(16) can be interpreted as the Euler-Lagrange equation of a
time-dependent density functional theory (TDDFT) \cite{p11}.
In Ref.~\cite{p11}, Eq.~(16) is approached via a variational scheme.
Here instead, Eq.~(16) is solved numerically by using a finite-difference
Crank-Nicolson predictor-corrector scheme \cite{p24}.
First we obtain the ground state by integrating Eq.~(16) in imaginary time.
Then we let a slightly perturbed wave function evolve in real time for
approximately one period of oscillation of the lowest (longitudinal)
frequency $\Omega_z$.
In the same time span, the density also undergoes several radial
oscillations of frequency $\Omega_\rho$.
We extract both frequencies by fitting the mean square widths of
$n({\bf r},t)$ with the sum of two cosines.
The breathing-mode frequencies obtained in this way are shown in Fig.~3 as
dots joined by solid lines.

The quantum pressure term is important for small number of atoms, as it
improves the determination of the density profile close to the surface of
the vapor cloud \cite{p11}.
For the number of particles of the experiments ($N\gtrsim10^5$), the
quantum pressure term is a relatively small correction, and according to
our calculation it accounts for about $0.5\%$ of the total energy.


Figure 3 shows substantial accord between PHA and TDDFT.
The main difference is the location of the predicted maximum in the bosonic
region $y>0$.
The differences are not only due to the approximations involved in the PHA,
but also to numerical errors in TDDFT introduced by space and time
discretization (estimated to less than 1\%)\cite{p25}.
We observe that, according to our predictions, the collective frequencies
reach the asymptotic large-$|y|$ limits more slowly than the theory of Hu
{\it et al.}\ \cite{p10} based on mean-field BCS Bogoliubov-de Gennes
equations within LDA.
In particular, in the BEC region, our numerical and analytical results 
(see the inset of Fig.~3) show that, contrary 
to the mean-field prediction, $\Omega_{\rho}$ approaches its asymptotic value
$2\,\omega_{\rho}$ passing through a local maximum. 
$\Omega_{z}$ has the same nonmonotonic behavior while reaching
$\sqrt{5/2}\,\omega_{z}$ for large $y$. 
This qualitative behavior was previoulsy suggested by Stringari \cite{p9}. 
The different asymptotic behavior of the BCS mean-field frequencies
is due to the neglect of beyond mean-field corrections.
Note also that further discrepancies on the BEC side are due to the
mean-field relation $a_m=2a_s$ rather than $a_m=0.6a_s$ as provided by
four-body scattering \cite{p16} and used in our calculation.
The $\Omega_\rho$ curve computed in Ref.~\cite{p11} (dot-dashed line in
Fig.~3) agrees rather well with our calculation.


Different theories are compared to the experimental data by Kinast {\it et
al.}\ \cite{p3} for the radial mode $\Omega_{\rho}$ and those by
Bartenstein {\it et al.}\ \cite{p4} for the longitudinal mode $\Omega_z$.
In Fig.~3, we use the standard variable $a_H/(N^{1/6} a) = 2^{1/2} 3^{1/6}
y$ and follow Ref.~\cite{p26} to determine the scattering length $a$ as a
function of the magnetic field $B$ near the Feshbach resonance:
\beq 
a = a_b  \left[ 1 + \alpha (B - B_0) \right] 
\left[ 1 + {\Delta \over B - B_0 } \right]  \; , 
\eeq 
where $B_0 = 83.4149$ mT, $a_b=-1405\;a_0$, $\Delta = 30.0$ mT, 
and $\alpha = 0.0040$ (mT)$^{-1}$. 
For the longitudinal frequency $\Omega_z$, our results are in
quantitative agreement with the experimental data, not unlike the
mean-field prediction.
The accord is less good for the radial mode. 
Experimental uncertainty of the position of the resonant field $B_0$ could
partly account for these discrepancies.
In particular the upward feature near $a_H/(N^{1/6} a)\simeq -1.5$ has been
related \cite{p3,p12} to the breaking of the Cooper pairs (due to sizeable
ratio between the collective energy $\hbar\Omega_{\rho}$ and the gap energy
$\Delta$ \cite{p12}), causing a failure of the hydrodynamical
approximation.
Finite-temperature and non-LDA effects not taken into account in the
theories could be relevant.
Note also that the experimental situation is not completely clear.
The experimental measurement of $\Omega_{\rho}$ performed by Bartenstein
{\it et al.}\ \cite{p4} (not shown in Fig.~3) disagree with the data of
Kinast {\it et al.}\ \cite{p3}.
In particular, Bartenstein {\it et al.}\ \cite{p4} find
$\Omega_{\rho}/\omega_{\rho}\simeq 1.6$ at the unitarity limit $y = 0$,
instead of the expected value $\Omega_{\rho}/\omega_{\rho} =
\sqrt{10/3}=1.82$ obtained from Eq.~(13) for $\gamma=2$, characteristic of
the (renormalized) free Fermi gas.

\section{Discussion}

We propose analytic expressions for the equations of state of a uniform
dilute Fermi gas across the BCS-BEC transition. 
These expressions are based on recent Monte-Carlo data and well-established
asymptotic expansions. 
By using a hydrodynamic local-density approximation we include the effect 
of harmonic confinement. We compare the predictions of this 
approach with experimental frequencies of
confined $^6$Li vapors. 
Other predicted physical quantites can be accessed by future experiments.
The hydrodynamic approach is improved, to address small number of
atoms, by including a quantum-pressure term.
Indeed, our parametric formula (6) provides an accurate expression for the
$\mu$ term in Eq.~(16) through Eq.~(7), which gives a better determination
of the density and of the collective frequencies.
This generalized quantum hydrodynamic approach, which takes into account
the beyond-mean-field corrections, provides a reliable tool to determine
the density profile of the Fermionic cloud and to investigate its
collective dynamical properties, including also mode coupling and
anharmonic oscillations.

\section*{Acknowledgement}

The authors thank A.\ Parola and L.\ Reatto for useful discussions, and
J.E.\ Thomas for drawing their attention to new experimental determinations
of $\Omega_\rho$.


\begin{thebibliography}{99}
\bibitem{p21} Our parametric function has been inspired by the 
simpler, but much less accurate, model function 
proposed by Combescot and Leyronars \cite{p12}.
\bibitem{p25} This accuracy is realized with a mesh of $200\times200$
radial$\times$longitudinal points, and a time step $\Delta t=6.25\cdot
10^{-5}/\omega_z$.
\end{thebibliography}
\end{document}